\documentclass[prb,twocolumn,showpacs]{revtex4}
\usepackage{graphicx} % \usepackage{bm} \usepackage{amsmath}

\usepackage{amsmath,amsthm,amssymb,mathrsfs}
\usepackage{bm}

\begin{document}

\title{Spin torque switching of an in-plane magnetized system in a thermally activated region}

\author{Tomohiro Taniguchi$^{1}$, Yasuhiro Utsumi$^{2}$, Michael Marthaler$^{3}$, Dmitri S. Golubev$^{4}$ and Hiroshi Imamura$^{1}$}
\email{h-imamura@aist.go.jp}
 \affiliation{
 $^{1}$Spintronics Research Center, National Institute of Advanced Industrial Science and Technology, Tsukuba, Ibaraki 305-8568, Japan, \\
 $^{2}$Faculty of Engineering, Mie University, Tsu, Mie, 514-8507, Japan, \\
 $^{3}$Institut f$\ddot{u}$r Theoretische Festk$\ddot{o}$rperphysik, Karlsruhe Institute of Technology, 76128 Karlsruhe, Germany, \\
 $^{4}$Institut f$\ddot{u}$r Nanotechnologie, Karlsruhe Institute of Technology, 76021 Karlsruhe, Germany.
 }

 \date{\today} 
 \begin{abstract}
  {
  The current dependence of the exponent of the spin torque switching rate 
  of an in-plane magnetized system was investigated 
  by solving the Fokker-Planck equation with low temperature and small damping and current approximations. 
  We derived the analytical expressions of the critical currents, $I_{\rm c}$ and $I_{\rm c}^{*}$. 
  At $I_{\rm c}$, the initial state parallel to the easy axis becomes unstable 
  while at $I_{\rm c}^{*} (\simeq 1.27 I_{\rm c})$ the switching occurs without the thermal fluctuation. 
  The current dependence of the exponent of the switching rate is well described by 
  $(1-I/I_{\rm c}^{*})^{b}$, 
  where the value of the exponent $b$ is approximately unity for $I \le I_{\rm c}$
  while $b$ rapidly increases up to $\sim 2.2$ with increasing current for $I_{\rm c} \le I \le I_{\rm c}^{*}$. 
  The linear dependence for $I \le I_{\rm c}$ agrees with the other works, 
  while the nonlinear dependence for $I_{\rm c} \le I \le I_{\rm c}^{*}$ was newly found by the present work. 
  The nonlinear dependence is important for analysis of the experimental results, 
  because most experiments are performed in the current region of $I_{\rm c} \le I \le I_{\rm c}^{*}$. 
  }
 \end{abstract}

 \maketitle

%====================================================================================================================================================%

%====================================================================================================================================================%

\section{Introduction}
\label{sec:Introduction}

%====================================================================================================================================================%

Spin torque switching \cite{slonczewski96,berger96} of a magnetization 
in nanostructured ferromagnets 
has attracted much attention 
due to its potential application to spintronics devices, 
such as spin random access memory (Spin RAM) \cite{yuasa08,suzuki08}. 
An accurate estimation of the thermal stability, $\Delta_{0}$, is very important 
for these applications, 
where the thermal stability is defined as 
the ratio of the magnetic anisotropy energy 
to the temperature ($k_{\rm B}T$). 
For example, 
the retention time of the Spin RAM exponentially depends on the thermal stability \cite{yakata09}. 
%and the linewidth of the power spectrum of the microwave oscillator is 
%inversely proportional to the thermal stability \cite{slavin09}. 

%====================================================================================================================================================%

The thermal stability is experimentally determined by 
measuring the spin torque switching in the thermally activated region \cite{yakata09,albert02}, 
and analyzing the switching probability  \cite{brown63} 
with the formula $P=1-\exp[-\int_{0}^{t} dt^{\prime}\nu(t^{\prime})]$, 
where $\nu=f \exp(-\Delta)$ and $f$ are 
the switching rate and the attempt frequency, respectively.  
Similar to many other non-equilibrium systems \cite{dykman80,dykman05,dykman08}, 
the exponent of the switching rate can be written in the form, 
\begin{equation}
  \Delta
  =
  \Delta_{0}
  \left(
    1
    -
    \frac{I}{I_{\rm c}}
  \right)^{b},
  \label{eq:barrier}
\end{equation}
is the energy barrier height of the switching divided by the temperature. 
In this paper, we call $\Delta$ the switching barrier. 
The thermal stability is defined as $\Delta_{0}$. 
Here, $I$ and $I_{\rm c}$ are 
the current and the critical current of the spin torque switching at zero temperature. 
Equation (\ref{eq:barrier}) characterizes the switching in the thermally activated region, 
and is applicable for $|I|<|I_{\rm c}|$. 
The exponent of the term $1-I/I_{\rm c}$ in Eq. (\ref{eq:barrier}) is denoted as $b$. 
Equation (\ref{eq:barrier}) was analytically derived for the uniaxially anisotropic system 
by solving the Fokker-Planck equation \cite{suzuki09,taniguchi11a,taniguchi12,butler12}, 
and $b=2$. 
Recently, it has also been confirmed by numerical simulations \cite{taniguchi12b,pinna12a}. 

%====================================================================================================================================================%

%====================================================================================================================================================%

%On the other hand, 
It is difficult to derive the analytical formula of the switching barrier 
for an in-plane magnetized system, 
which does not have axial symmetry due to the presence of two anisotropic axes 
(an easy axis along the in-plane and a hard axis normal to the plane). 
Previous analyses adopted $b=1$, 
which has been obtained by solving the Landau-Lifshitz-Gilbert (LLG) equation \cite{koch04} 
as well as the Fokker-Planck equation \cite{apalkov05}. 
The point of these works is that 
the effect of the spin torque on the switching barrier can be described by 
the effective temperature \cite{koch04} $T/(1-I/I_{\rm c})$. 
However, the effective temperature approach is valid only for $I \ll I_{\rm c}$ \cite{koch04}. 
The current dependence of the switching barrier in a relatively large current region $(I \simeq I_{\rm c})$ 
remains unclear,
while such large current is applied 
to a ferromagnet in experiments to quickly observe the switching. 
These facts motivated us to study the switching barrier in the large current region. 
In Ref. \cite{taniguchi12b}, 
we numerically solved the LLG equation of the in-plane magnetized system, 
and found that the switching does not occur even if $I > I_{\rm c}$, 
although $I_{\rm c}$ has been assumed to be the critical current 
of the spin torque switching at zero temperature. 
We also found that the exponent $b$ is larger than 2 
by analyzing the numerical results with a phenomenological model of the switching \cite{taniguchi12d,taniguchi11d,taniguchi12a}.

%====================================================================================================================================================%

%====================================================================================================================================================%

In this paper, 
we develop an analytical theory of the spin torque switching 
of an in-plane magnetized system 
based on the Fokker-Planck theory with WKB approximations \cite{benjacob82,maier93,smelyanskiy97,sukhorukov07,dykman80,dykman05,dykman08,kamenev11}, 
where the temperature is assumed to be low, 
and the magnitudes of the damping and the spin torque are assumed to be small. 
By assuming that the solution of the Fokker-Planck equation is given by $W \propto \exp(-\Delta)$, 
we find that the switching barrier is given by 
\begin{equation}
  \Delta
  =
  -V
  \int_{E_{\rm min}}^{E_{\rm max}} dE 
  \frac{H_{\rm s}F_{\rm s}-\alpha MF_{\alpha}}{\alpha k_{\rm B}T MF_{\alpha}}.
  \label{eq:barrier_definition_intro}
\end{equation} 
The physical meaning of Eq. (\ref{eq:barrier_definition_intro}) is as follows. 
The numerator in the integral arises from 
the drift terms of the Fokker-Planck equation, 
where the terms $H_{\rm s}F_{\rm s}$ and $\alpha MF_{\alpha}$ are proportional to 
the work done by spin torque and the energy dissipation due to the damping, respectively. 
On the other hand, 
the denominator arises from the diffusion term of the Fokker-Planck equation, 
where the thermal fluctuation depends on the damping constant $\alpha$ and the temperature $T$ 
through the fluctuation-dissipation theorem. 
Then, the switching barrier is given by the integral of the ratio 
between the drift and the diffusion terms. 
The boundaries of the integral, $E_{\rm min}$ and $E_{\rm max}$, are defined 
by the region of $H_{\rm s}F_{\rm s}<\alpha MF_{\alpha}$, 
because the energy dissipation due to the damping overcomes the spin torque in this region, 
and thus, the energy absorption from the thermal bath is required 
to climb up the switching barrier. 
The thermally activated region is defined as the current region $I \le I_{\rm c}^{*}$, 
where $I_{\rm c}^{*}$ satisfies $H_{\rm s}F_{\rm s}= \alpha MF_{\alpha}$ 
at the saddle point of the energy map. 
The relation between $I_{\rm c}$ in the conventional theory \cite{sun00} and $I_{\rm c}^{*}$ is as follows. 
The initial state parallel to the easy axis becomes unstable 
at $I=I_{\rm c}$. 
However, the condition $I>I_{\rm c}$ does not guarantee the switching. 
On the other hand, at $I=I_{\rm c}^{*}$, 
the switching occurs without the thermal fluctuation. 
We derive the analytical expression of $I_{\rm c}^{*}$, 
and find that $I_{\rm c}^{*}\simeq 1.27 I_{\rm c}$. 
We also find that the current dependence of the switching barrier is well described by 
$(1-I/I_{\rm c}^{*})^{b}$, 
where the value of the exponent $b$ is approximately unity for the current $I \le I_{\rm c}$, 
while $b$ rapidly increases up to $\sim 2.2$ with increasing current for $I_{\rm c} \le I \le I_{\rm c}^{*}$, 
showing a good agreement with Refs. \cite{koch04,apalkov05}. 
The nonlinear dependence for $I_{\rm c} \le I \le I_{\rm c}^{*}$ newly found in this paper is important
to evaluate the thermal stability 
because most experiments are performed in the current region of $I_{\rm c} \le I \le I_{\rm c}^{*}$. 

%====================================================================================================================================================%

The paper is organized as follows. 
In Sec. \ref{sec:Fokker-Planck_Equation_for_Magnetization_Switching}, 
we summarize the Fokker-Planck equation of the magnetization switching. 
In Sec. \ref{sec:WKB_Approximation} 
we describe the details of the WKB approximation. 
Equation (\ref{eq:barrier_definition}), 
(\ref{eq:lambda_E}) and (\ref{eq:energy_change}) leads to the main results in next sections. 
The readers who are interested in the applications of Eq. (\ref{eq:barrier_definition_intro}) can 
directly move to the next sections, 
Secs. \ref{sec:Uniaxially_Anisotropic_System} and \ref{sec:In-plane_Magnetized_System}. 
In Sec. \ref{sec:Uniaxially_Anisotropic_System}, 
we apply the above formula to the uniaxially anisotropic system, 
and show that the present formula reproduces the results $b=2$ 
derived in Refs. \cite{suzuki09,taniguchi12,butler12}. 
In Sec. \ref{sec:In-plane_Magnetized_System}, 
the switching barrier of an in-plane magnetized system is calculated. 
In Sec. \ref{sec:Comparison_with_Previous_Works}, 
we compare the current results with our previous works. 
Section \ref{sec:Summary} is devoted to the conclusion. 

%====================================================================================================================================================%

%====================================================================================================================================================%

\section{Fokker-Planck Equation for Magnetization Switching}
\label{sec:Fokker-Planck_Equation_for_Magnetization_Switching}

%====================================================================================================================================================%

The LLG equation is given by \cite{landau35,landau80,gilbert04}
\begin{equation}
  \frac{d \mathbf{m}}{d t}
  =
  -\gamma
  \mathbf{m}
  \times
  \mathbf{H}
  -
  \gamma
  H_{\rm s}
  \mathbf{m}
  \times
  \left(
    \mathbf{n}_{\rm p}
    \times
    \mathbf{m}
  \right)
  +
  \alpha
  \mathbf{m}
  \times
  \frac{d \mathbf{m}}{d t},
  \label{eq:LLG}
\end{equation}
where 
the gyromagnetic ratio and the Gilbert damping constant are denoted as $\gamma$ and $\alpha$, respectively. 
$\mathbf{m}=(\sin\theta\cos\varphi,\sin\theta\sin\varphi,\cos\theta)$ and $\mathbf{n}_{\rm p}$ are 
the unit vectors pointing to the directions of the magnetization of 
the free and pinned layers, respectively. 
$\mathbf{H}=-\partial E/ \partial (M \mathbf{m})$ is the magnetic field, 
where $M$ and $E$ are 
the saturation magnetization 
and the magnetic energy density. 
For an in-plane magnetized system, $E$ is given by 
\begin{equation}
\begin{split}
  E
  &=
  -\frac{MH_{\rm K}}{2}
  \left(
    \mathbf{m}
    \cdot
    \mathbf{e}_{z}
  \right)^{2}
  +
  2 \pi M^{2}
  \left(
    \mathbf{m}
    \cdot
    \mathbf{e}_{x}
  \right)^{2}
\\
  &=
  -\frac{MH_{\rm K}}{2}
  z^{2}
  +
  2\pi M^{2}
  \left(
    1 - z^{2}
  \right)
  \cos^{2}\varphi,
  \label{eq:energy}
\end{split}
\end{equation}
where $z=\cos\theta$, and 
$H_{\rm K}$ and $4\pi M$ are 
the uniaxial anisotropy field along the easy ($z$) axis 
and the demagnetization along the hard ($x$) axis, respectively. 
For the uniaxially anisotropic system discussed in Sec. \ref{sec:Uniaxially_Anisotropic_System}, 
the demagnetization field should be absent. 
In Secs. \ref{sec:Uniaxially_Anisotropic_System} and \ref{sec:In-plane_Magnetized_System}, 
we assume that $\mathbf{n}_{\rm p}=\mathbf{e}_{z}$. 
The strength of the spin torque in the unit of the magnetic field, 
$H_{\rm s}$, is given by 
\begin{equation}
  H_{\rm s}
  =
  \frac{\hbar \eta I}{2eMV},
\end{equation}
where $\eta$ and $V$ are the spin polarization of the current 
and the volume of the free layer, respectively. 
Although the explicit form of the energy density is specified in Eq. (\ref{eq:energy}), 
the extension of the following formula to the general system is straightforward.

Let us express Eq. (\ref{eq:LLG}) in terms of the canonical variables 
because the following formula is based on the canonical theory 
developed in Refs. \cite{benjacob82,maier93,smelyanskiy97,sukhorukov07,kamenev11}. 
The magnetization dynamics without the spin torque and dissipation is described by 
the following Lagrangian density \cite{braun96,maekawa06}, 
\begin{equation}
  L
  =
  -\frac{M}{\gamma}
  \dot{\varphi}
  \left(
    \cos\theta
    -
    1
  \right)
  -
  E.
  \label{eq:Lagrangian_conserved}
\end{equation}
The first term of Eq. (\ref{eq:Lagrangian_conserved}) 
is the solid angle of the magnetization dynamics in the spin space, 
or equivalently, the Berry phase. 
The canonical coordinate is $q=\varphi$, 
and the momentum is defined as $p=\partial L/\partial \dot{q}=-(M/\gamma)(\cos\theta-1)$. 
In terms of the canonical variables $(q,p)$, 
the LLG equation (\ref{eq:LLG}) of the uniaxial and the in-plane magnetized system 
can be expressed as 
\begin{equation}
\begin{split}
  \frac{d q}{d t}
  =&
  \frac{1}{1+\alpha^{2}}
  \frac{\partial E}{\partial p}
  \!-\!
  \!\frac{\alpha\gamma g^{-1}}{(1+\alpha^{2})M}
  \frac{\partial E}{\partial q}
  \!+\!
  \frac{\alpha\gamma H_{\rm s}}{1+\alpha^{2}},
  \label{eq:LLG_q}
\end{split}
\end{equation}
\begin{equation}
\begin{split}
  \frac{d p}{d t}
  =&
  -\!\frac{1}{1+\alpha^{2}}
  \frac{\partial E}{\partial q}
  \!-\!
  \!\frac{\alpha gM }{(1+\alpha^{2})\gamma}
  \frac{\partial E}{\partial p}
  \!+\!
  \frac{gMH_{\rm s}}{1+\alpha^{2}},
  \label{eq:LLG_p}
\end{split}
\end{equation}
where $g=\sin^{2}\theta$. 
It should be noted that 
the spin torque term cannot be expressed as 
a gradient of the potential energy $E$, 
in general, 
and can be regarded as a damping or anti-damping. 

%====================================================================================================================================================%

At a finite temperature, 
the random torque, $-\gamma \mathbf{m}\times \mathbf{h}$, should be added to 
the right-hand side of Eq. (\ref{eq:LLG}), 
where the components of the random field $\mathbf{h}$ satisfy the fluctuation-dissipation theorem \cite{brown63}, 
\begin{equation}
  \langle h_{i}(t) h_{j}(t^{\prime}) \rangle 
  =
  \frac{2D}{\gamma^{2}}
  \delta_{ij}
  \delta(t-t^{\prime}).
  \label{eq:FDT}
\end{equation}
Here $D=\alpha \gamma k_{\rm B}T/(MV)$ is the diffusion coefficient. 
Due to the random torque, 
the magnetization switching can be regarded as 
the two-dimensional Brownian motion of a point particle in 
the $(q,p)$ phase space. 

%====================================================================================================================================================%

Let us introduce the distribution function of the magnetization, $W$, 
in the $(q,p)$ phase space. 
The Fokker-Planck equation is given by \cite{brown63}
\begin{equation}
\begin{split}
  \frac{\partial W}{\partial t}
  =&
  -\frac{\partial}{\partial q}
  \frac{dq}{d t}
  W
  +
  \frac{D}{1+\alpha^{2}}
  \frac{\partial}{\partial q}
  g^{-1}
  \frac{\partial}{\partial q}
  W
\\
  &-
  \frac{\partial}{\partial p}
  \frac{d p}{d t}
  W
  +
  \frac{D}{1+\alpha^{2}}
  \left(
    \frac{M}{\gamma}
  \right)^{2}
  \frac{\partial}{\partial p}
  g
  \frac{\partial}{\partial p}
  W,
  \label{eq:Fokker-Planck}
\end{split}
\end{equation}
where $dq/dt$ and $dp/dt$ should be regarded as 
the right-hand sides of Eqs. (\ref{eq:LLG_q}) and (\ref{eq:LLG_p}). 
The terms proportional to $D$ correspond to the diffusion terms 
while the others correspond to the drift terms. 
In equilibrium ($\partial W/\partial t=0$ and $H_{\rm s}=0$), 
the distribution function is identical to the Boltzmann distribution function 
($\propto \exp[-E/(k_{\rm B}T)]$).

%====================================================================================================================================================%

\section{WKB Approximation}
\label{sec:WKB_Approximation}

%====================================================================================================================================================%

In general, 
the distribution function determined by Eq. (\ref{eq:Fokker-Planck}) depends 
on the two variables, $(q,p)$, 
and it is very difficult to solve Eq. (\ref{eq:Fokker-Planck}) for an arbitrary system. 
Thus, we use the following two assumptions. 

%====================================================================================================================================================%

First, the low temperature assumption 
corresponding to the WKB approximations in Refs. \cite{benjacob82,maier93,smelyanskiy97,sukhorukov07,dykman80,dykman05,dykman08,kamenev11} is employed. 
We assume that the distribution function takes the following form \cite{benjacob82,smelyanskiy97}, 
\begin{equation}
  W 
  \propto 
  \exp
  \left(
    -\alpha S/D
  \right). 
\end{equation}
In the zero temperature limit ($\alpha/D \ll 1$), 
Eq. (\ref{eq:Fokker-Planck}) is reduced to \cite{comment1}
\begin{equation}
\begin{split}
  \frac{\partial S}{\partial t}
  =&
  -\frac{dq}{d t}
  \frac{\partial S}{\partial q}
  -
  \alpha 
  g^{-1}
  \left(
    \frac{\partial S}{\partial q}
  \right)^{2}
\\
  &
  -\frac{d p}{d t}
  \frac{\partial S}{\partial p}
  -
  \alpha g
  \left(
    \frac{M}{\gamma}
  \right)^{2}
  \left(
    \frac{\partial S}{\partial p}
  \right)^{2}. 
\end{split}
\end{equation}
Here $S$ and $-\partial S/\partial t \equiv \mathscr{H}$ can be regarded as 
the effective action and the Hamiltonian density, respectively \cite{smelyanskiy97,sukhorukov07,kamenev11}. 
The corresponding Lagrangian density is then given by 
\begin{equation}
  \mathscr{L}
  =
  -\dot{q}
  \lambda_{q}
  -
  \dot{p}
  \lambda_{p}
  -
  \mathscr{H}, 
  \label{eq:Lagrangian_definition}
\end{equation}
where $\lambda_{q}=-\partial S/\partial q$ and $\lambda_{p}=-\partial S/\partial p$ 
conjugated to $q$ and $p$ are the counting variables \cite{sukhorukov07}. 
The effective action is given by $S=\int dt \mathscr{L}$. 

%====================================================================================================================================================%

Second, we utilize the fact that 
the Gilbert damping constant and the spin torque strength are small. 
The values of the Gilbert damping constant $\alpha$ for the conventional ferromagnetic materials \cite{oogane06}
such as Co, Fe, and Ni are on the order of $10^{-2}$. 
Also, the critical current of the spin torque switching \cite{sun00}
is proportional to $\alpha |\mathbf{H}|$. 
Then, the switching time of the magnetization is much longer than 
the precession period $\tau$, 
and the energy $E$ is approximately conserved 
during one period of the precession. 
Following Ref. \cite{sukhorukov07}, 
we introduce the new canonical variables $(E,s)$ 
accompanied by the new counting variables $(\lambda_{E},\lambda_{s})$. 
Then, the Lagrangian density is expressed as 
[see Appendix \ref{sec:Appendix_A}]
\begin{equation}
  \mathscr{L}
  =
  -\frac{dE}{dt}
  \lambda_{E}
  -
  \mathscr{H}_{E},
  \label{eq:Lagrangian}
\end{equation}
\begin{equation}
\begin{split}
  \mathscr{H}_{E}
  =&
  \lambda_{E}
  \frac{\partial E}{\partial q}
  \!\!
  \left[
    \frac{\alpha MH_{\rm s}}{1+\alpha^{2}}
    \frac{\partial}{\partial p}
    \mathbf{m}
    \!\cdot\!
    \mathbf{n}_{\rm p}
    \!+\!
    \frac{\alpha\gamma}{(1+\alpha^{2})M}
    g^{-1}
    \frac{\partial E}{\partial q}
  \right.
\\
  &\ \ \ \ \ \ \ \ \ \ \ \ \ 
  \left.
    +
    \frac{\gamma H_{\rm s}}{1+\alpha^{2}}
    g^{-1}
    \frac{\partial}{\partial q}
    \mathbf{m}
    \!\cdot\!
    \mathbf{n}_{\rm p}
  \right]
\\
  &-
  \lambda_{E}
  \frac{\partial E}{\partial p}
  \!\!
  \left[
    \frac{\alpha MH_{\rm s}}{1+\alpha^{2}}
    \frac{\partial}{\partial q}
    \mathbf{m}
    \!\cdot\!
    \mathbf{n}_{\rm p}
    \!-\!
    \frac{\alpha M}{(1+\alpha^{2})\gamma}
    g
    \frac{\partial E}{\partial p}
  \right.
\\
  &\ \ \ \ \ \ \ \ \ \ \ \ \ 
  \left.
    -\frac{M^{2}H_{\rm s}}{(1+\alpha^{2})\gamma}
    g
    \frac{\partial}{\partial p}
    \mathbf{m}
    \!\cdot\!
    \mathbf{n}_{\rm p}
  \right]
\\
  &+
  \lambda_{E}^{2}
  \frac{\alpha}{1+\alpha^{2}}
  g^{-1}
  \left(
    \frac{\partial E}{\partial q}
  \right)^{2}
\\
  &+
  \lambda_{E}^{2}
  \frac{\alpha}{1+\alpha^{2}}
  g
  \left(
    \frac{M}{\gamma}
  \right)^{2}
  \left(
    \frac{\partial E}{\partial p}
  \right)^{2}. 
  \label{eq:Hamiltonian}
\end{split}
\end{equation}

%====================================================================================================================================================%

We perform the time average of $\mathscr{L}$ 
over the constant energy orbit, 
$\overline{\mathscr{L}}=\int_{0}^{\tau} dt \mathscr{L}/\tau = -[M/(\gamma\tau)]\int_{0}^{2\pi} d\varphi \mathscr{L}/(\partial E/ \partial z)$. 
Here we use the fact that the constant orbit is determined by 
the non-perturbative equation of motion, 
$d\varphi/dt=-(\gamma/M)(\partial E/\partial z)$. 
The averaged Lagrangian density is expressed as 
\begin{equation}
  \overline{\mathscr{L}}
  =
  -\frac{\gamma}{M}
  \frac{dE}{dt}
  \lambda_{E}
  -
  \overline{\mathscr{H}}_{E},
\end{equation}
where 
we renormalize $\lambda_{E}$ as $(M \lambda_{E}/\gamma) \to \lambda_{E}$ 
by which $\lambda_{E}$ becomes a dimensionless variable. 
The effective Hamiltonian is given by 
\begin{equation}
\begin{split}
  \overline{\mathscr{H}}_{E}
  =&
  -\lambda_{E}
  \frac{\gamma H_{\rm s}}{1+\alpha^{2}}
  F_{\rm s}
  +
  \lambda_{E}
  (1+\lambda_{E})
  \frac{\alpha\gamma M}{1+\alpha^{2}}
  F_{\alpha}, 
  \label{eq:Hamiltonian_uniaxial}
\end{split}
\end{equation}
where $F_{\rm s}$ and $F_{\alpha}$ are given by 
\begin{equation}
  F_{\rm s}
  =
  \frac{1}{\tau}
  \!
  \int_{0}^{2\pi} 
  \!\!
  \frac{d\varphi}{\partial E/\partial z}
  \left[
    (1-z^{2})
    \left(
      \frac{\partial E}{\partial z}
    \right)
    \!-\!
    \alpha
    \left(
      \frac{\partial E}{\partial\varphi}
    \right)
  \right],
  \label{eq:F_s}
\end{equation}
\begin{equation}
  F_{\alpha}
  =
  -\frac{1}{\tau M^{2}}
  \!
  \int_{0}^{2\pi}
  \!\!
  \frac{d\varphi}{\partial E/\partial z}
  \left[
    (1-z^{2})
    \left(
      \frac{\partial E}{\partial z}
    \right)^{2}
    \!+\!
    \frac{1}{1-z^{2}}
    \left(
      \frac{\partial E}{\partial \varphi}
    \right)^{2}
  \right].
  \label{eq:F_alpha}
\end{equation}
The term proportional to $H_{\rm s}$ in Eq. (\ref{eq:Hamiltonian_uniaxial}) describes the work done by the spin torque, 
while the term proportional to $\alpha \lambda_{E}$ in the second term of Eq. (\ref{eq:Hamiltonian_uniaxial}) 
describes the energy dissipation due to the Gilbert damping [Appendix \ref{sec:Appendix_B}]. 
The term proportional to $\alpha \lambda_{E}^{2}$ 
arises from the thermal fluctuation. 

%====================================================================================================================================================%

The switching barrier is obtained by integrating the Lagrangian density 
along the switching path. 
It is sufficient to calculate the integral along the optimal path \cite{kamenev11} 
in the low temperature limit, 
where the optimal path corresponds to the switching path 
obtained by Eq. (\ref{eq:LLG}) without the thermal fluctuation. 
The optimal path \cite{kamenev11} can be found by solving $\overline{\mathscr{H}}_{E}=0$. 
One of the solutions is given by $E=-MH_{\rm K}/2$, 
where $z=\pm 1$ and $\varphi$ is arbitrary, 
due to which $F_{\rm s}=F_{\alpha}=0$. 
The other two solutions of $\overline{\mathscr{H}}_{E}=0$ are given by 
$\lambda_{E}=0,\lambda_{E}^{*}$, 
where $\lambda_{E}^{*}$ is given by 
\begin{equation}
  \lambda_{E}^{*}
  =
  -1
  +
  \frac{H_{\rm s}F_{\rm s}}{\alpha MF_{\alpha}}.
  \label{eq:lambda_E}
\end{equation}
The equation of motion along $\lambda_{E}=0$ describes the drift in the energy space 
due to the competition between the spin torque and the Gilbert damping at zero temperature. 
The energy change $\dot{E}=-\partial \overline{\mathscr{H}}_{E}/\partial \lambda_{E}$ along $\lambda_{E}=0$ is given by 
\begin{equation}
  \dot{E}
  \big|_{\lambda_{E}=0}
  =
  \frac{\gamma (H_{\rm s}F_{\rm s} - \alpha M F_{\alpha})}{1+\alpha^{2}}.
  \label{eq:energy_change}
\end{equation}
On the other hand, $\lambda_{E}^{*}$ corresponds to 
the time reversal path of $\lambda_{E}=0$ \cite{kamenev11}, 
and thus, $\dot{E}|_{\lambda_{E}=\lambda_{E}^{*}}=-\dot{E}|_{\lambda_{E}=0}$. 
The switching barrier, $\Delta=\alpha S/D$, is then given by  
\begin{equation}
  \Delta
  =
  -\frac{V}{k_{\rm B}T}
  \int_{E_{\rm min}}^{E_{\rm max}} dE 
  \lambda_{E}^{*},
  \label{eq:barrier_definition}
\end{equation} 
which is identical to Eq. (\ref{eq:barrier_definition_intro}). 
The boundaries of the integral in Eq. (\ref{eq:barrier_definition}) are determined 
as follows. 
According to its definition, Eq. (\ref{eq:lambda_E}), 
$\lambda_{E}^{*}$ depends on 
the work done by spin torque and the energy dissipation due to the damping, 
see also Appendix \ref{sec:Appendix_B}. 
In the region $\lambda_{E}^{*}>0$, 
the spin torque overcomes the damping, 
and the magnetization can move from the initial state parallel to the easy axis 
without thermal fluctuation. 
On the other hand,  
the damping exceeds the spin torque in the region $\lambda_{E}^{*}<0$,
and thus, the thermal fluctuation is required to achieve the switching. 
Then, the integral in Eq. (\ref{eq:barrier_definition}) in the region $\lambda_{E}^{*}<0$ 
gives the switching barrier, 
i.e., the boundaries of the integral in Eq. (\ref{eq:barrier_definition}) are 
those of $\lambda_{E}^{*}<0$. 
In the latter sections, examples of $\lambda_{E}^{*}$ are shown
[see Figs. \ref{fig:fig1} and \ref{fig:fig5}]. 
It should be noted that the condition $\lambda_{E}^{*}<0$ is identical to 
$H_{\rm s}F_{\rm s}<\alpha MF_{\alpha}$ discussed after Eq. (\ref{eq:barrier_definition_intro}) 
because the energy dissipation due to the damping ($\propto -F_{\alpha}$) is always negative. 

%====================================================================================================================================================%

Equation (\ref{eq:barrier_definition}) with Eqs. (\ref{eq:lambda_E}) and (\ref{eq:energy_change}) is 
the main result in this section, 
and enables us to calculate the current dependence of the switching barrier 
in the next sections. 
Equation (\ref{eq:barrier_definition}) is similar to Eq. (19) of Ref. \cite{apalkov05}, 
except for the additional condition on the integral range 
determined by Eq. (\ref{eq:energy_change}). 
We emphasize that this difference is crucial to obtain the exponent $b$, 
as shown in Secs. \ref{sec:Uniaxially_Anisotropic_System} and \ref{sec:In-plane_Magnetized_System}.

%====================================================================================================================================================%

%====================================================================================================================================================%

%====================================================================================================================================================%

\section{Uniaxially Anisotropic System}
\label{sec:Uniaxially_Anisotropic_System}

%====================================================================================================================================================%

%====================================================================================================================================================%

\begin{figure}%[p]
\centerline{\includegraphics[width=1.0\columnwidth]{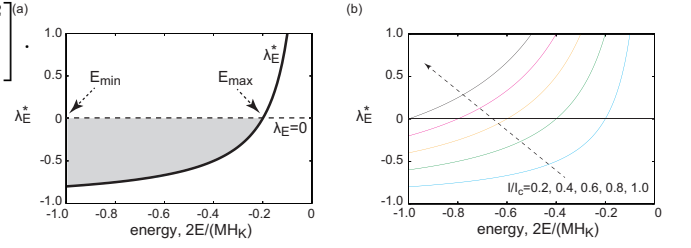}}\vspace{-3.0ex}
\caption{
         (a) A typical dependence of $\lambda_{E}^{*}$ on the energy $E$ for the uniaxially anisotropic system. 
             The energy is normalized by the factor $MH_{\rm K}/2$. 
             The dotted line corresponds to $\lambda_{E}=0$. 
             The switching barrier is obtained by integrating $\lambda_{E}^{*}$ from $E_{\rm min}$ to $E_{\rm max}$, 
             i.e., the shaded region. 
             The lower boundary $E_{\rm min}$ is fixed to $-MH_{\rm K}/2$, 
             while the upper boundary $E_{\rm max}$ located at $-(MH_{\rm K}/2)(I/I_{\rm c})^{2}$. 
         (b) The dependence of $\lambda_{E}^{*}$ for the current $I/I_{\rm c}=0.2,0.4,0.6,0.8$ and $1.0$. 
         \vspace{-3ex}}
\label{fig:fig1}
\end{figure}

%====================================================================================================================================================%

%====================================================================================================================================================%

%====================================================================================================================================================%

In this section, 
we calculate the switching barrier of the uniaxially anisotropic system, 
in which the energy density is given by $E=-(MH_{\rm K}/2)\cos^{2}\theta$. 
Here $z=\cos\theta$ can be expressed in terms of $E$ and $H_{\rm K}$ as 
$z=\sqrt{-2E/(MH_{\rm K})}$. 
The metastable states of the magnetization locate at $\mathbf{m}=\pm \mathbf{e}_{z}$, 
and the initial state is taken to be $\mathbf{m}=\mathbf{e}_{z}$. 
The functions $F_{\rm s}$ and $F_{\alpha}$ (Eqs. (\ref{eq:F_s}) and (\ref{eq:F_alpha})) of this system are given by 
\begin{equation}
  F_{\rm s}
  =
  \gamma
  H_{\rm K}
  z
  \left(
    1
    -
    z^{2}
  \right),
\end{equation}
\begin{equation}
  F_{\alpha}
  =
  \frac{\gamma H_{\rm K}^{2}}{M}
  z^{2}
  \left(
    1
    -
    z^{2}
  \right),
\end{equation}
respectively.
The critical current $I_{\rm c}$ of the uniaxially anisotropic system is given by 
\begin{equation}
  I_{\rm c}
  =
  \frac{2 \alpha e MV}{\hbar \eta}
  H_{\rm K}.
\end{equation}

%====================================================================================================================================================%

%====================================================================================================================================================%

The switching path, $\lambda_{E}^{*}$ in Eq. (\ref{eq:lambda_E}), 
is given by $\lambda_{E}^{*}=-1+[I/(I_{\rm c}z)]$. 
Figure \ref{fig:fig1} (a) shows a typical dependence of $\lambda_{E}^{*}$ on the energy $E$. 
The region $\lambda_{E}^{*}<0$ is from $E_{\rm min}=-MH_{\rm K}/2$ to $E_{\rm max}=-(MH_{\rm K}/2)(I/I_{\rm c})^{2}$. 
Figure \ref{fig:fig1} (b) shows $\lambda_{E}^{*}$ for various currents. 
The upper boundary of $\lambda_{E}^{*}<0$, $E_{\rm max}$, is zero at $I=0$, 
and approaches to $-MH_{\rm K}/2$ with increasing current. 
At $I=I_{\rm c}$, $E_{\rm max}$ equals $E_{\rm min}$. 
This behavior is particular for the uniaxially anisotropic system, 
where the switching can be reduced to the one-dimensional problem 
due to the rotational symmetry along the $z$ axis. 
In this case, the effect of the spin torque can be described by 
the effective potential \cite{suzuki09,taniguchi11a} $E_{\rm eff}=-(MH_{\rm K}/2)\cos^{2}\theta+(MH_{\rm s}/\alpha)\cos\theta$. 
The effective potential has two minima at $\theta=0,\pi$ and one maximum at $\theta=\cos^{-1}(I/I_{\rm c})$. 

%====================================================================================================================================================%

%====================================================================================================================================================%

\begin{figure}%[p]
\centerline{\includegraphics[width=0.7\columnwidth]{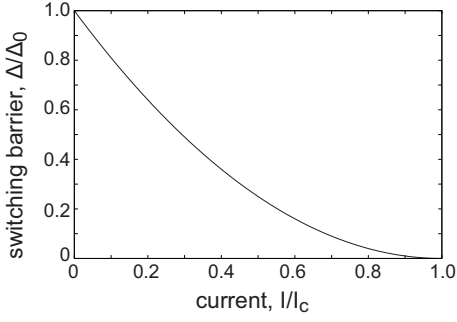}}\vspace{-3.0ex}
\caption{
         The dependence of the switching barrier of the uniaxially anisotropic system on the current $I/I_{\rm c}$. 
         The barrier height is normalized by $\Delta_{0}$. 
         \vspace{-3ex}}
\label{fig:fig2}
\end{figure}

%====================================================================================================================================================%

%====================================================================================================================================================%

%====================================================================================================================================================%

The switching barrier, Eq. (\ref{eq:barrier_definition}), is obtained by 
integrating the region of $\lambda_{E}^{*}<0$, 
i.e., the shaded region of Fig. \ref{fig:fig1} (a), 
and is given by 
\begin{equation}
  \Delta
  =
  \Delta_{0}
  \left(
    1
    -
    \frac{I}{I_{\rm c}}
  \right)^{2},
\end{equation}
where $\Delta_{0}=MH_{\rm K}/(2k_{\rm B}T)$ is the thermal stability. 
Figure \ref{fig:fig2} shows the dependence of the switching barrier 
on the current $I/I_{\rm c}$. 
The barrier height is normalized by $\Delta_{0}$. 
The exponent of the term $1-I/I_{\rm c}$ is $b=2$ in this system, 
which is consistent with Refs. \cite{suzuki09,taniguchi11a}. 

%====================================================================================================================================================%

%====================================================================================================================================================%

\section{In-plane Magnetized System}
\label{sec:In-plane_Magnetized_System}

\begin{figure}%[p]
\centerline{\includegraphics[width=1.0\columnwidth]{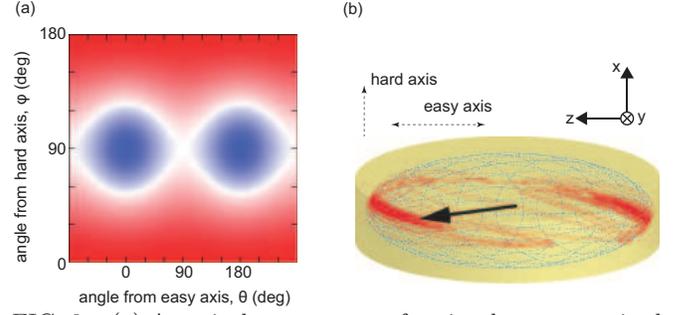}}\vspace{-3.0ex}
\caption{
         (a) A typical energy map of an in-plane magnetized system. 
             There are two minima $(E=-MH_{\rm K}/2)$ at $\theta=0,\pi$,
             two maxima $(E=2\pi M^{2})$ at $(\theta,\varphi)=(\pi/2,0),(\pi/2,\pi)$, 
             and two saddle points $(E=0)$ at $(\theta,\varphi)=(\pi/2,\pi/2),(\pi/2,3\pi/2)$. 
             The blue and red regions correspond to the stable $(E<0)$ and unstable $(E>0)$ regions, 
             respectively. 
         (b) A typical switching orbit in the in-plane magnetized system. 
             The value of the parameters are $M=1000$ emu/c.c., $H_{\rm K}=200$ Oe, $\alpha=0.01$, and $T=20$ K, 
             see Ref. \cite{taniguchi12b}. 
         \vspace{-3ex}}
\label{fig:fig3}
\end{figure}

In this section, 
we investigate the switching barrier of an in-plane magnetized system. 
The energy density of this system is given by Eq. (\ref{eq:energy}). 
The magnitudes of the demagnetization and the uniaxial anisotropy fields 
of the ferromagnetic materials of conventional in-plane Spin RAM are 
on the order of 1 T and 100 Oe \cite{yakata09}, respectively. 
Thus, %in almost all cases we are interested in, 
$H_{\rm K}/(4\pi M)$ is on the order of $10^{-2}$. 
Figure \ref{fig:fig3} (a) shows a typical energy map of an in-plane magnetized system. 
There are two minima of the energy ($E=-MH_{\rm K}/2$) at $\theta=0,\pi$ 
and two maxima ($E=2\pi M^{2}$) at $(\theta,\varphi)=(\pi/2,0),(\pi/2,\pi)$. 
Because of the large demagnetization field, 
the magnetization switches through one of the saddle points ($E=0$) 
at $(\theta,\varphi)=(\pi/2,\pi/2),(\pi/2,3\pi/2)$. 
Figure \ref{fig:fig3} (b) shows a typical switching orbit 
obtained by numerically solving the LLG equation \cite{taniguchi12b}. 
Starting from $\mathbf{m}=\mathbf{e}_{z}$, 
the magnetization precesses around the easy axis, 
and gradually approaches the saddle point. 
Then, the magnetization passes close to the saddle point, 
and relaxes to the other stable state, $\mathbf{m}=-\mathbf{e}_{z}$. 
Since the switching time is dominated by 
the time spent during the precession around the easy axis 
in which $-MH_{\rm K}/2 \le E \le 0$, 
it is sufficient to evaluate $\lambda_{E}^{*}$ of Eq. (\ref{eq:barrier_definition}) 
in this energy region. 

%====================================================================================================================================================%

%====================================================================================================================================================%

%====================================================================================================================================================%

%====================================================================================================================================================%

%====================================================================================================================================================%

The details of the calculation of the switching barrier are as follows. 
The variable $z$ relates to $E$, $H_{\rm K}$, and $4\pi M$ as 
\begin{equation}
  z
  =
  \sqrt{
    \frac{4\pi M \cos^{2}\varphi-2E/M}{4\pi M \cos^{2}\varphi+H_{\rm K}}
  }. 
\end{equation}
Then, the functions $F_{\rm s}$ and $F_{\alpha}$ (Eqs. (\ref{eq:F_s}) and (\ref{eq:F_alpha})) are given by 
\begin{equation}
\begin{split}
  F_{\rm s}
  &=
  \frac{2\pi (H_{\rm K} \!+\! 2E/M)}{\tau \sqrt{H_{\rm K}(H_{\rm K} \!+\! 4\pi M)}},
  \label{eq:F1_in_plane}
\end{split}
\end{equation}
\begin{equation}
\begin{split}
  F_{\alpha}
  &=
  \frac{4}{\tau M}
  \sqrt{
    \frac{4\pi M \!-\! 2E/M}{H_{\rm K}}
  }
\\
  &\ \ \times
  \left[
    \frac{2E}{M}
    \mathsf{K}
    \!
    \left(
      \!\sqrt{
        \frac{4\pi M(H_{\rm K} \!+\! 2E/M)}{H_{\rm K}(4\pi M \!-\! 2E/M)}
      }
    \right)
  \right.
\\
  &\ \ \ \ 
  \left.
    +
    H_{\rm K}
    \mathsf{E}
    \!
    \left(
      \!\sqrt{
        \frac{4\pi M(H_{\rm K} \!+\! 2E/M)}{H_{\rm K}(4\pi M \!-\! 2E/M)}
      }
    \right)
  \right].
  \label{eq:F2_in_plane}
\end{split}
\end{equation}
where the precession period is given by 
\begin{equation}
  \tau
  =
  \frac{4}{\gamma \sqrt{H_{\rm K}(4\pi M \!-\! 2E/M)}}
  \mathsf{K}
  \!
  \left(
    \!\sqrt{
      \frac{4\pi M(H_{\rm K} \!+\! 2E/M)}{H_{\rm K}(4\pi M \!-\! 2E/M)}
    }
  \right).
  \label{eq:period_in_plane}
\end{equation}
The complete elliptic integrals of the first and second kinds are defined as 
$\mathsf{K}(k) \!=\! \int_{0}^{1} dy /\sqrt{(1 \!-\! k^{2}y^{2})(1 \!-\! y^{2})}$ and 
$\mathsf{E}(k) \!=\! \int_{0}^{1} dy \sqrt{(1 \!-\! k^{2}y^{2})/(1 \!-\! y^{2})}$, respectively. 
In the limit of $4\pi M \to 0$, 
$\tau$, $F_{\rm s}$, and $F_{\alpha}$ are identical to 
those calculated for the uniaxially anisotropic system. 

%====================================================================================================================================================%

%====================================================================================================================================================%

\begin{figure}%[p]
\centerline{\includegraphics[width=0.8\columnwidth]{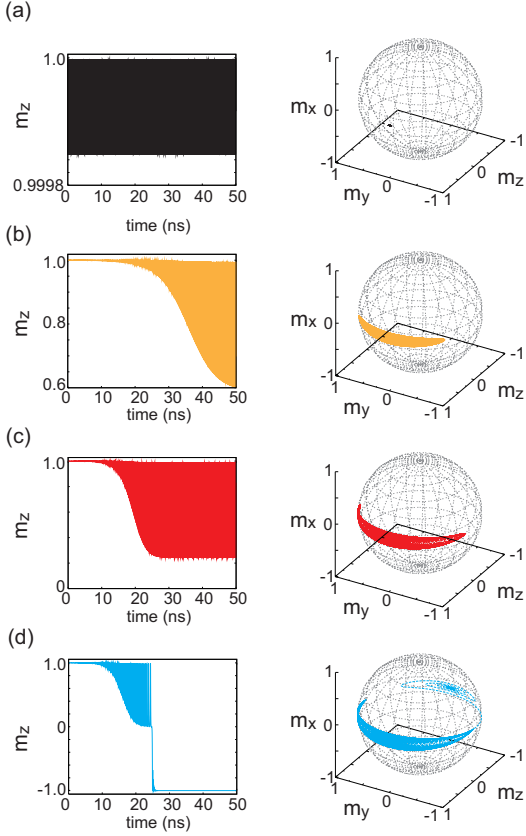}}\vspace{-3.0ex}
\caption{
         The magnetization dynamics at zero temperature. 
         The time evolution of $m_{z}$ (left) and the dynamic orbit (right) are shown. 
         (a) At $I=I_{\rm c}$. 
             The magnetization oscillates around the initial state ($\mathbf{m}=\mathbf{e}_{z}$) with small amplitude. 
         (b), (c) At $I=r(I_{\rm c}^{*}-I_{\rm c})+I_{\rm c}$ ($r=0.4$ for (b) and $r=0.8$ for (c)). 
             The oscillation amplitude increases with increasing current. 
         (d) At $I=I_{\rm c}^{*}$, the switching occurs. 
         \vspace{-3ex}}
\label{fig:fig4}
\end{figure}

%====================================================================================================================================================%

%====================================================================================================================================================%

It should be noted that $\lambda_{E}^{*}$ of Eq. (\ref{eq:lambda_E}) satisfies the following relations: 
\begin{equation}
  \lim_{E \to -MH_{\rm K}/2}
  \lambda_{E}^{*}
  =
  -\left(
    1
    -
    \frac{I}{I_{\rm c}}
  \right),
  \label{eq:limit_lambda_E_min}
\end{equation}
\begin{equation}
  \lim_{E \to 0}
  \lambda_{E}^{*}
  =
  -\left(
    1
    -
    \frac{I}{I_{\rm c}^{*}}
  \right),
  \label{eq:limit_lambda_E_max}
\end{equation}
where the critical currents $I_{\rm c}$ and $I_{\rm c}^{*}$ are, 
respectively, given by 
\begin{equation}
  I_{\rm c}
  =
  \frac{2 \alpha e MV}{\hbar \eta}
  \left(
    H_{\rm K}
    +
    2 \pi M
  \right).
  \label{eq:critical_current}
\end{equation}
\begin{equation}
  I_{\rm c}^{*}
  =
  \frac{4\alpha eMV}{\pi \hbar \eta}
  \sqrt{
    4\pi M 
    \left(
      H_{\rm K}
      +
      4\pi M
    \right)
  }.
  \label{eq:switching_current}
\end{equation}
Below, we concentrate on the region of $I_{\rm c}<I_{\rm c}^{*}$ ($H_{\rm K}/(4\pi M) < 0.196$). 
In the conventional ferromagnetic thin film, 
$H_{\rm K}$ and $4\pi M$ are on the order of $100$ Oe and $1$ T, 
and thus, this condition is usually satisfied. 
For an infinite demagnetization field limit,
$I_{\rm c}^{*}/I_{\rm c}\simeq 4/\pi=1.27$...

%====================================================================================================================================================%

%====================================================================================================================================================%

\begin{figure}%[p]
\centerline{\includegraphics[width=1.0\columnwidth]{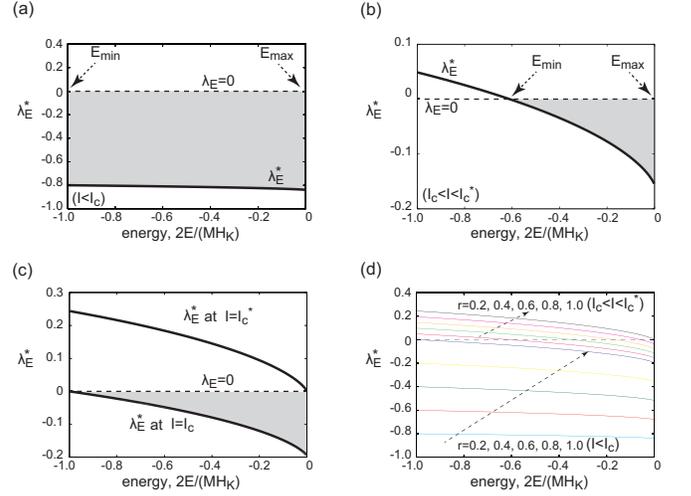}}\vspace{-3.0ex}
\caption{
         (a) Typical dependence of $\lambda_{E}^{*}$ of the in-plane magnetized system on the energy $E$ for $I<I_{\rm c}$. 
             The switching barrier is obtained by integrating $\lambda_{E}^{*}$ 
             from $E_{\rm min}=-MH_{\rm K}/2$ to $E_{\rm max}=0$, i.e., the shaded region. 
         (b) Typical dependence of $\lambda_{E}^{*}$ on the energy for $I_{\rm c}<I<I_{\rm c}^{*}$. 
             The lower boundary of the integral, $E_{\rm min}$, locates at $-MH_{\rm K}/2<E_{\rm min}<0$. 
         (c) $\lambda_{E}^{*}$ for $I=I_{\rm c}$ and $I=I_{\rm c}^{*}$. 
         (d) The dependence of $\lambda_{E}^{*}$ on the various currents, 
             $I=rI_{\rm c}$ for $I \le I_{\rm c}$ and $I=r(I_{\rm c}^{*}-I_{\rm c})+I_{\rm c}$ for $I_{\rm c}<I \le I_{\rm c}^{*}$, 
             where $r=0.2,0.4,0.6,0.8$ and $1.0$. 
         \vspace{-3ex}}
\label{fig:fig5}
\end{figure}

%====================================================================================================================================================%

%====================================================================================================================================================%

%====================================================================================================================================================%

The physical meanings of $I_{\rm c}$ and $I_{\rm c}^{*}$ are as follows. 
In Fig. 4, we show the magnetization dynamics at the zero temperature. 
The parameters are $M=1000$ emu/c.c., $H_{\rm K}=200$ Oe, $\gamma=17.64$ MHz/Oe, $\alpha=0.01$, 
$\eta=0.8$, and $V=\pi \times 80 \times 35 \times 2.5$ nm${}^{3}$, respectively \cite{taniguchi12b}, 
by which $I_{\rm c}=0.54$ mA and $I_{\rm c}^{*}=0.68$ mA, respectively. 
For $I<I_{\rm c}$, the damping overcomes the spin torque, 
and the initial state parallel to the easy axis is stable. 
At $I=I_{\rm c}$, the initial state becomes unstable, 
and the magnetization oscillates around the easy axis with a small constant amplitude 
[see Fig. \ref{fig:fig4} (a)]. 
This has been already pointed out in Ref. \cite{sun00}, 
and $I_{\rm c}$ has been considered to be the critical current of the magnetization switching. 
However, we emphasize that $I=I_{\rm c}$ is not a critical point, 
because $I=I_{\rm c}$ does not guarantee the switching \cite{taniguchi12b}. 
At $I_{\rm c}<I<I_{\rm c}^{*}$, 
the oscillation amplitude increases with increasing current, 
as shown in Figs. \ref{fig:fig4} (b) and (c). 
The critical current is $I_{\rm c}^{*}$, 
over which the magnetization can switch its direction 
without the thermal fluctuation, 
as shown in Fig. \ref{fig:fig4} (d).

%====================================================================================================================================================%

In Fig. \ref{fig:fig5}, 
we show the dependence of $\lambda_{E}^{*}$ on the energy for various currents. 
A typical $\lambda_{E}^{*}$ for $I<I_{\rm c}$ is shown in Fig. \ref{fig:fig5} (a). 
Here, $\lambda_{E}^{*}$ is always negative 
because the damping exceeds the spin torque. 
The switching barrier is obtained by integrating $\lambda_{E}^{*}$ 
from $E_{\rm min}=-MH_{\rm K}/2$ to $E_{\rm max}=0$. 
On the other hand, Fig. \ref{fig:fig5} (b) shows a typical $\lambda_{E}^{*}$ for $I_{\rm c}<I<I_{\rm c}^{*}$. 
From the initial state ($E=-MH_{\rm K}/2$) to a certain energy $E_{\rm min}$, 
the magnetization can move without the thermal fluctuation 
because the spin torque overcomes the damping ($\lambda_{E}^{*}>0$). 
At $E=E_{\rm min}$, the magnetization dynamics is on a stable orbit, 
where the spin torque and the damping are balanced. 
The thermal fluctuation is required from $E_{\rm min}$ to $E_{\rm max}$ to switch the magnetization direction. 
In Fig. \ref{fig:fig5} (c), 
we show $\lambda_{E}^{*}$ for $I=I_{\rm c}$ and $I=I_{\rm c}^{*}$. 
At $I=I_{\rm c}$, $\lambda_{E}^{*} \le 0$, and $\lambda_{E}^{*}$ is zero at $E=-MH_{\rm K}/2$. 
On the other hand, at $I=I_{\rm c}^{*}$, 
$\lambda_{E}^{*} \ge 0$, and $\lambda_{E}^{*}=0$ at $E=0$. 
Figure \ref{fig:fig5} (d) shows $\lambda_{E}^{*}$ for various currents. 

%====================================================================================================================================================%

%====================================================================================================================================================%
 %====================================================================================================================================================%

%====================================================================================================================================================%

\begin{figure}%[p]
\centerline{\includegraphics[width=0.7\columnwidth]{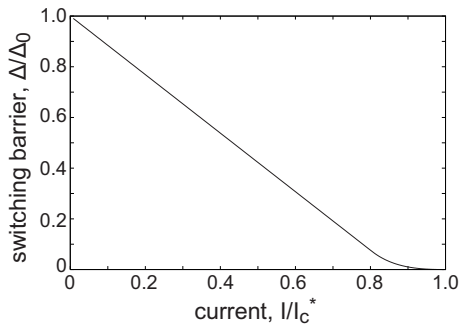}}\vspace{-3.0ex}
\caption{
         The current dependence of the switching barrier normalized by $\Delta_{0}$, 
         i.e., $\Delta/\Delta_{0}$. 
         The current magnitude is normalized by $I_{\rm c}^{*}$. 
         \vspace{-3ex}}
\label{fig:fig6}
\end{figure}

%====================================================================================================================================================%

%====================================================================================================================================================%

\begin{figure}%[p]
\centerline{\includegraphics[width=0.7\columnwidth]{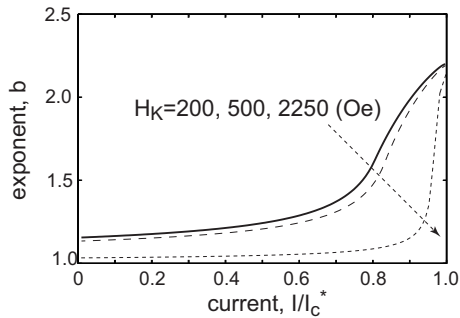}}\vspace{-3.0ex}
\caption{
         The estimated value of the exponent $b$ 
         by analyzing Fig. \ref{fig:fig6} with a function $\ln(\Delta/\Delta_{0})/\ln(1-I/I_{\rm c}^{*})$. 
         The values of $H_{\rm K}$ in the solid, dashed, and dotted lines are 
         200, 500, and 2250 Oe, respectively. 
         \vspace{-3ex}}
\label{fig:fig7}
\end{figure}

Figure \ref{fig:fig6} shows the current dependence of the switching barrier 
obtained by numerically integrating the $\lambda_{E}^{*}$ shown in Fig. \ref{fig:fig5} (d)
[see Eq. (\ref{eq:barrier_definition})], 
in which the barrier height is normalized by $\Delta_{0}$. 
The upper boundary of the integral range, $E_{\rm max}$, is taken to be $E=0$. 
On the other hand, the lower boundary, $E_{\rm min}$, is $-MH_{\rm K}/2$ for $I \le I_{\rm c}$, 
while it is determined by numerically solving $\lambda_{E}^{*}=0$ for $I_{\rm c} < I \le I_{\rm c}^{*}$. 
We find that the current dependence of the switching barrier is well described by $(1-I/I_{\rm c}^{*})^{b}$ 
[see also Appendix \ref{sec:Appendix_C}]. 
The dependence is approximately linear ($b \simeq 1$) for $I \le I_{\rm c}$ 
showing the consistence with Refs. \cite{koch04,apalkov05}. 
On the other hand, 
we find a nonlinear dependence for $I_{\rm c} < I \le I_{\rm c}^{*}$. 
The solid line in Fig. \ref{fig:fig7} shows the dependence of the exponent $b$ on the current, 
where $b$ is determined by the switching barrier shown in Fig. \ref{fig:fig6} as 
$b \equiv \ln (\Delta/\Delta_{0})/\ln(1-I/I_{\rm c}^{*})$. 
As shown, $b$ slightly increases with increasing current for $I < I_{\rm c} (\simeq 0.8 I_{\rm c}^{*})$.
For $I_{\rm c} \le I \le I_{\rm c}^{*}$, 
$b$ rapidly increases, and reaches $b>2$ near $I \lesssim I_{\rm c}^{*}$. 
It should also be noted that $b$ is not universal. 
The values of $b$ with different $H_{\rm K}$ values are 
also shown in Fig. \ref{fig:fig7} 
by the dashed ($H_{\rm K}=500$ Oe) 
and dotted ($H_{\rm K}=2250 {\rm \ Oe} \simeq 0.18 \times 4\pi M$) lines. 
As shown, the value of $b$ decreases with increasing $H_{\rm K}$. 
A further increase of $H_{\rm K}$ breaks the condition ($I_{\rm c} < I_{\rm c}^{*}$, or equivalently, $H_{\rm K}/(4\pi M) <0.196$), 
and beyond the scope of this paper. 

%====================================================================================================================================================%

The nonlinear dependence of the switching barrier is important 
for an accurate estimation of the thermal stability. 
Typical experiments are performed in the large current region, $I \lesssim I_{\rm c}^{*}$, 
to quickly measure the switching. 
By applying a linear fit to Fig. \ref{fig:fig7}, 
as done in the analysis of experiments \cite{yakata09,albert02}, 
the thermal stability would be significantly underestimated. 

%====================================================================================================================================================%

%====================================================================================================================================================%

%====================================================================================================================================================%

%====================================================================================================================================================%

%====================================================================================================================================================%

%====================================================================================================================================================%

\section{Comparison with Previous Works}
\label{sec:Comparison_with_Previous_Works}

%====================================================================================================================================================%

In Sec. \ref{sec:WKB_Approximation}, 
we assume that the switching time from the initial state ($E=-MH_{\rm K}/2$) 
to the saddle point ($E=0$) 
is much longer than the precession period $\tau$. 
Since the precession period (Eq. (\ref{eq:period_in_plane})) diverges at $E=0$, 
this assumption is apparently violated in the vicinity of the saddle point. 
The condition under which our approximation is valid is 
\begin{equation}
  \tau 
  |\dot{E}|
  \big|_{E=0}
  \ll
  \frac{MH_{\rm K}}{2}.
  \label{eq:condition_validity}
\end{equation}
By using Eqs. (\ref{eq:energy_change}), (\ref{eq:F1_in_plane}), and (\ref{eq:F2_in_plane}), 
we find that Eq. (\ref{eq:condition_validity}) can be expressed as 
\begin{equation}
  4\pi M 
  \ll
  \frac{H_{\rm K}}{16 \alpha^{2}}. 
  \label{eq:condition_approx}
\end{equation}
The parameters used in Sec. \ref{sec:In-plane_Magnetized_System} satisfy this condition. 

%====================================================================================================================================================%

According to Eq. (\ref{eq:condition_approx}), 
the present formula does not work for an infinite demagnetization field limit, 
$4\pi M/H_{\rm K} \to \infty$, 
where the switching occurs completely in-plane without precession. 
Then, the switching barrier is given by 
$\Delta_{0}(1-I/I_{\rm c})^{2}$ 
with $b=2$ and $I_{\rm c}=2 \alpha eMV H_{\rm K}/(\hbar \eta)$, 
as shown in our previous work, \cite{taniguchi11a}. 
Thus, the present work is valid for Eq. (\ref{eq:condition_approx}), 
while the previous work is valid for $4\pi M/H_{\rm K} \to \infty$. 
The switching barrier in the intermediate region, 
$H_{\rm K}/(16 \alpha^{2}) \lesssim 4\pi M $, remains unclear, 
and is beyond the scope of this paper. 

%====================================================================================================================================================%

Let us also discuss the relation between the present work and one of our previous works in Ref. \cite{taniguchi12b}. 
In Ref. \cite{taniguchi12b}, 
we estimate that $b \sim 3$ for the in-plane magnetized system 
by numerically solving the LLG equation at various temperatures 
and adopting the phenomenological model of the switching \cite{taniguchi12d}. 
The parameters are identical to those used in Fig. \ref{fig:fig7}. 
The switching currents at $0.1$ K and $20$ K are 
$0.67$ mA ($\simeq 0.99 I_{\rm c}^{*}$) and $0.62$ mA ($\simeq 0.92 I_{\rm c}^{*}$), respectively \cite{comment3}. 

%====================================================================================================================================================%

%====================================================================================================================================================%

%====================================================================================================================================================%

\begin{figure}%[p]
\centerline{\includegraphics[width=0.7\columnwidth]{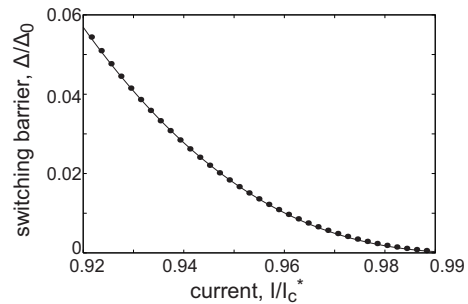}}\vspace{-3.0ex}
\caption{
         The dots are the current dependence of the switching barrier 
         in $0.92 I_{\rm c}^{*} \le I \le 0.99 I_{\rm c}^{*}$. 
         The solid line represents a fitting proportional to $(1-I/I_{\rm c}^{*})^{b^{\prime}}$, 
         where $b^{\prime}$ is assumed to be constant in this current region. 
         \vspace{-3ex}}
\label{fig:fig8}
\end{figure}

%====================================================================================================================================================%

%====================================================================================================================================================%

The main difference between the present work and Refs. \cite{taniguchi12b,taniguchi12d} is that, 
in Refs. \cite{taniguchi12b,taniguchi12d} 
the exponent $b$ is assumed to be constant. 
In Fig. \ref{fig:fig8}, 
the switching barrier shown in Fig. \ref{fig:fig6} is fitted by a function proportional to $(1-I/I_{\rm c}^{*})^{b^{\prime}}$, 
with a constant $b^{\prime}$. 
The current range is from $0.92 I_{\rm c}^{*}$ to $0.99 I_{\rm c}^{*}$, 
according to Ref. \cite{taniguchi12b}. 
The obtained value of $b^{\prime}$ is 2.5, 
which is close to the estimated value ($b \sim 3$) in Ref. \cite{taniguchi12b}. 
The difference may arise from the temperature dependence of $\Delta$, or 
the current dependence of the attempt frequency neglected in the present work. 

%====================================================================================================================================================%

%====================================================================================================================================================%

%====================================================================================================================================================%

\section{Summary}
\label{sec:Summary}

In summary, 
we have developed a theory of the spin torque switching 
of the in-plane magnetized system 
based on the Fokker-Planck theory with WKB approximation. 
We derived the analytical expressions of the critical currents, $I_{\rm c}$ and $I_{\rm c}^{*}$. 
The initial state parallel to the easy axis becomes unstable 
at $I=I_{\rm c}$, 
which has been derived in Ref. \cite{sun00}. 
On the other hand, at $I=I_{\rm c}^{*} (\simeq 1.27 I_{\rm c})$, 
the switching occurs without the thermal fluctuation. 
We also find that the current dependence of the switching barrier is well described by 
$(1-I/I_{\rm c}^{*})^{b}$, 
where the value of the exponent $b$ is approximately linear for the current $I \le I_{\rm c}$, 
while $b$ rapidly increases with increasing current for $I_{\rm c} \le I \le I_{\rm c}^{*}$. 
The nonlinear dependence for $I_{\rm c} \le I \le I_{\rm c}^{*}$ is important 
for an accurate evaluation of the thermal stability 
because most experiments are performed in the current region of $I_{\rm c} \le I \le I_{\rm c}^{*}$.

%====================================================================================================================================================%

%====================================================================================================================================================%

\section*{Acknowledgement}

The authors would like to acknowledge 
H. Kubota, H. Maehara, A. Emura, A. Fukushima, K. Yakushiji, T. Yorozu, H. Arai, K. Ando, S. Yuasa, S. Miwa, Y. Suzuki, H. Sukegawa and S. Mitani 
for their valuable discussions. 
This work was supported by JSPS KAKENHI Number 23226001. 

\textit{Note}
After we had submitted the manuscript, 
D. Pinna, A. D. Kent, and D. L. Stein informed us that 
they worked in a similar problem independently. 

%====================================================================================================================================================%

%====================================================================================================================================================%

%====================================================================================================================================================%

\appendix
\section{Derivation of Eqs. (\ref{eq:Lagrangian}) and (\ref{eq:Hamiltonian})}
\label{sec:Appendix_A}

First, we divide the Lagrangian density $\mathscr{L}$, Eq. (\ref{eq:Lagrangian_definition}), 
into non-perturbative ($\mathscr{L}_{0}$) and perturbative ($\mathscr{L}_{1}$) parts. 
In terms of the canonical variables $(q,p)$, 
Eq. (\ref{eq:LLG}) can be expressed as 
\begin{equation}
\begin{split}
  \frac{d q}{d t}
  =&
  \frac{1}{1+\alpha^{2}}
  \frac{\partial E}{\partial p}
  \!-\!
  \frac{\alpha MH_{\rm s}}{1+\alpha^{2}}
  \frac{\partial}{\partial p}
  \mathbf{m}
  \!\cdot\!
  \mathbf{n}_{\rm p}
\\
  &-
  \!\frac{\alpha\gamma}{(1+\alpha^{2})M}
  g^{-1}
  \frac{\partial E}{\partial q}
  \!-\!
  \frac{\gamma H_{\rm s}}{1+\alpha^{2}}
  g^{-1}
  \frac{\partial}{\partial q}
  \mathbf{m}
  \!\cdot\!
  \mathbf{n}_{\rm p},
  \label{eq:LLG_q_gen}
\end{split}
\end{equation}
\begin{equation}
\begin{split}
  \frac{d p}{d t}
  =&
  -\!\frac{1}{1+\alpha^{2}}
  \frac{\partial E}{\partial q}
  \!+\!
  \frac{\alpha MH_{\rm s}}{1+\alpha^{2}}
  \frac{\partial}{\partial q}
  \mathbf{m}
  \cdot
  \mathbf{n}_{\rm p}
\\
  &-
  \!\frac{\alpha M}{(1+\alpha^{2})\gamma}
  g
  \frac{\partial E}{\partial p}
  \!-\!
  \frac{M^{2}H_{\rm s}}{(1+\alpha^{2})\gamma}
  g
  \frac{\partial}{\partial p}
  \mathbf{m}
  \!\cdot\!
  \mathbf{n}_{\rm p}.
  \label{eq:LLG_p_gen}
\end{split}
\end{equation}
Equations (\ref{eq:LLG_q_gen}) and (\ref{eq:LLG_p_gen}) can be directly obtained from Eq. (\ref{eq:LLG}) 
for general system 
by expressing Eq. (\ref{eq:LLG}) in terms of the spherical coordinate, $(\theta,\varphi)$. 

By using the explicit forms of $dq/dt$ and $dp/dt$ in $\mathscr{H}$ (Eqs. (\ref{eq:LLG_q_gen}) and (\ref{eq:LLG_p_gen})), 
the non-perturbative Lagrangian density is given by 
\begin{equation}
\begin{split}
  \mathscr{L}_{0}
  =&
  -\lambda_{q}
  \!
  \left(
    \frac{dq}{dt}
    \!-\!
    \frac{1}{1+\alpha^{2}}
    \frac{\partial E}{\partial p}
  \right)
  \!-\!
  \lambda_{p}
  \!
  \left(
    \frac{dp}{dt}
    \!+\!
    \frac{1}{1+\alpha^{2}}
    \frac{\partial E}{\partial q}
  \right).
\end{split}
\end{equation}
On the other hand, 
the perturbative Lagrangian density is given by 
\begin{equation}
\begin{split}
  \mathscr{L}_{1}
  =&
  -\lambda_{q}
  \!
  \left[
    \frac{\alpha MH_{\rm s}}{1+\alpha^{2}}
    \frac{\partial}{\partial p}
    \mathbf{m}
    \!\cdot\!
    \mathbf{n}_{\rm p}
    +
    \frac{\alpha\gamma}{(1+\alpha^{2})M}
    g^{-1}
    \frac{\partial E}{\partial q}
  \right.
\\
  &\ \ \ \ \ \ 
  \left.
    +
    \frac{\gamma H_{\rm s}}{1+\alpha^{2}}
    g^{-1}
    \frac{\partial}{\partial q}
    \mathbf{m}
    \!\cdot\!
    \mathbf{n}_{\rm p}
  \right]
\\
  &+
  \lambda_{p}
  \!
  \left[
    \frac{\alpha MH_{\rm s}}{1+\alpha^{2}}
    \frac{\partial}{\partial q}
    \mathbf{m}
    \!\cdot\!
    \mathbf{n}_{\rm p}
    -
    \frac{\alpha M}{(1+\alpha^{2})\gamma}
    g
    \frac{\partial E}{\partial p}
  \right.
\\
  &\ \ \ \ \ \ 
  \left.
    -
    \frac{M^{2}H_{\rm s}}{(1+\alpha^{2})\gamma}
    g
    \frac{\partial}{\partial p}
    \mathbf{m}
    \!\cdot\!
    \mathbf{n}_{\rm p}
  \right]
\\
  &-
  \lambda_{q}^{2}
  \frac{\alpha}{1+\alpha^{2}}
  g^{-1}
  -
  \lambda_{p}^{2}
  \frac{\alpha}{1+\alpha^{2}}
  g
  \left(
    \frac{M}{\gamma}
  \right)^{2}. 
\end{split}
\end{equation}

The canonical transformation from $(q,p)$ to $(E,s)$ 
is accompanied by the canonical transformation of the counting variables as 
$\lambda_{q}=(\partial E/\partial q)\lambda_{E}+(\partial s/\partial q)\lambda_{s}$ 
and $\lambda_{p}=(\partial E/\partial p)\lambda_{E}+(\partial s/\partial p)\lambda_{s}$. 
The non-perturbative Lagrangian is then given by 
$\mathscr{L}_{0}=-\lambda_{E}(dE/dt)-\lambda_{s}[(ds/dt)-[s,H]_{\rm P}]$, 
where $d/dt=(\partial /\partial q)(dq/dt)+(\partial/\partial p)(dp/dt)$ and 
$[,]_{\rm P}$ is the Poisson bracket. 
In the small damping limit, $s$ can be regarded as a physical time $t$. 
Since $s$ is a conjugated variable of $E$ ($[s,H]_{\rm P}=1$), 
the non-perturbative Lagrangian density is given by Eq. (\ref{eq:Lagrangian}). 
Since the switching barrier is determined by $\mathscr{L}_{0}$ and is independent of $\lambda_{s}$, 
we set $\lambda_{s}=0$, 
according to Ref. \cite{sukhorukov07}. 
Then, $\mathscr{H}_{E}=-\mathscr{L}_{1}$ is given by Eq. (\ref{eq:Hamiltonian}). 

%====================================================================================================================================================%

%\appendix
\section{Work done by spin torque and damping}
\label{sec:Appendix_B}

Here we show that the functions $F_{\rm s}$ and $F_{\alpha}$ are proportional 
to the work done by the spin torque and damping. 

The time evolution of the magnetic energy,
$dE/dt=-M\mathbf{H}\cdot(d\mathbf{m}/dt)$, 
is calculated by using Eq. (\ref{eq:LLG}) as 
\begin{equation}
  \frac{dE}{dt}
  =
  \mathcal{W}_{\rm s}
  +
  \mathcal{W}_{\alpha},
\end{equation}
where $\mathcal{W}_{\rm s}$ and $\mathcal{W}_{\alpha}$ are given by 
\begin{equation}
  \mathcal{W}_{\rm s}
  =
  \frac{\gamma MH_{\rm s}}{1+\alpha^{2}} 
  \left[
    \mathbf{n}_{\rm p}
    \cdot
    \mathbf{H}
    \!-\!
    \left(
      \mathbf{m}
      \cdot
      \mathbf{n}_{\rm p}
    \right)
    \left(
      \mathbf{m}
      \cdot
      \mathbf{H}
    \right)
    \!-\!
    \alpha 
    \mathbf{n}_{\rm p}
    \cdot
    \left(
      \mathbf{m}
      \times
      \mathbf{H}
    \right)
  \right],
\end{equation}
\begin{equation}
  \mathcal{W}_{\alpha}
  =
  -\frac{\alpha\gamma M}{1+\alpha^{2}}
  \left[
    \mathbf{H}^{2}
    -
    \left(
      \mathbf{m}
      \cdot
      \mathbf{H}
    \right)^{2}
  \right]. 
\end{equation}
Here, $\mathcal{W}_{\rm s}$ is the work done by spin torque 
while $\mathcal{W}_{\alpha}$ is the energy dissipation due to the damping. 
It should be noted that $\mathcal{W}_{\alpha}$ is always negative, 
while $\mathcal{W}_{\rm s}$ is either positive or negative 
depending on the current direction. 
Let us consider the in-plane magnetized system as an example, 
where the magnetic field is given by 
$\mathbf{H}=(-4\pi M m_{x},0,H_{\rm K}m_{z})$. 
Then, the time averages of $\mathcal{W}_{\rm s}$ and $\mathcal{W}_{\alpha}$ over one period 
of the precession around the easy axis are given by 
\begin{equation}
  \overline{\mathcal{W}}_{\rm s}
  =
  \frac{MH_{\rm s}}{1+\alpha^{2}}
  F_{\rm s},
\end{equation}
\begin{equation}
  \overline{\mathcal{W}}_{\alpha}
  =
  -\frac{\alpha M^{2}}{1+\alpha^{2}}
  F_{\alpha},
\end{equation}
respectively, 
where $F_{\rm s}$ and $F_{\alpha}$ are given by Eqs. (\ref{eq:F1_in_plane}) and (\ref{eq:F2_in_plane}), respectively. 
Thus, we can verify that $F_{\rm s}$ and $F_{\alpha}$ are proportional to 
the work done by the spin torque and the damping, respectively. 

%====================================================================================================================================================%

%\appendix
\section{Current dependence of the switching barrier near $I_{\rm c}^{*}$}
\label{sec:Appendix_C}

Here, we derive an analytical formula of the switching barrier near $I_{\rm c}^{*}$. 
For simplicity, 
we use the normalized variables 
$k=H_{\rm K}/4\pi M$, $\varepsilon=E/4\pi M^{2}$, and $s=H_{\rm s}/4\pi M$. 
First, let us derive the approximated form of $\lambda_{E}$ (Eq. (\ref{eq:lambda_E})) 
around $\varepsilon \sim 0$. 
The Taylor expansions of $H_{\rm s}F_{\rm s}\tau$ and $\alpha M F_{\alpha} \tau$ are respectively given by 
\begin{equation}
  \frac{H_{\rm s}F_{\rm s}\tau}{4\pi M}
  =
  \frac{2\pi(k+2 \varepsilon)s}{\sqrt{k(1+k)}},
\end{equation}
\begin{equation}
  \frac{\alpha MF_{\alpha}\tau}{4\pi M}
  \simeq
  \alpha
  \left(
    4 \sqrt{k}
    +
    \frac{2(1-k)\varepsilon}{\sqrt{k}}
    \left\{
      1
      -
      \ln
      \left[
        -\frac{(1+k)\varepsilon}{8k}
      \right]
    \right\}
  \right), 
\end{equation}
where $\varepsilon \ln \varepsilon$ term appears and thus it is non-analytic at $\epsilon=0$. 
Then, the approximated $\lambda_{E}^{*}$ is given by 
\begin{equation}
\begin{split}
  \lambda_{E}^{*}
  \simeq
  &
  -\left(
    1
    -
    \frac{I}{I_{\rm c}^{*}}
  \right)
\\
  &+
  \frac{I \varepsilon}{2 I_{\rm c}^{*} k}
  \left\{
    3
    +
    k
    +
    \left(
      1 - k
    \right)
    \ln
    \left[
      -\frac{(1+k) \varepsilon}{8 k}
    \right]
  \right\}.
\end{split}
\end{equation}
The energy $E_{\rm min}=4\pi M^{2}\varepsilon_{0}$ corresponding to the intersection of $\lambda_{E}^{*}$ and $\lambda_{E}=0$ can be obtained 
by solving the following self-consistent relation: 
\begin{equation}
  \varepsilon_{0}
  =
  \left(
    1
    -
    \frac{I}{I_{\rm c}^{*}}
  \right)
  \frac{2k(I_{\rm c}^{*}/I)}{3+k+(1-k)\ln[-(1+k)\varepsilon_{0}/(8k)]}.
  \label{eq:energy_approximated}
\end{equation}
Up to the first order of $E_{\rm min}$, 
the switching barrier is given by 
\begin{equation}
  \Delta
  \simeq
  -\frac{E_{\rm min}V}{k_{\rm B}T}
  \left(
    1
    -
    \frac{I}{I_{\rm c}^{*}}
  \right)
  \label{eq:barrier_approximated}
\end{equation}
where $E_{\rm min}=4\pi M^{2} \varepsilon_{0}$ is determined by Eq. (\ref{eq:energy_approximated}). 
Equation (\ref{eq:barrier_approximated}) is only valid close to the switching current $I_{\rm c}^{*}$ where $E_{\rm min}$ locates near $E=0$. 

%====================================================================================================================================================%

%\bibliography{biblist}% Produces the bibliography via BibTeX.

\begin{thebibliography}{35}
\bibitem[{com({\natexlab{a}})}]{comment1}
\bibinfo{note}{In our definition, the dimension of both $S$ and $D$ are taken
  to be 1/time, to make the notation simple. The dimension of $S$ can be
  changed to that of the action by multiplying the factor $V/\gamma^{2}$.}
\bibitem[{com({\natexlab{b}})}]{comment3}
\bibinfo{note}{The switching time shown in Fig. 4 of Ref. \cite{taniguchi12b}
  is converted to the switching current by multiplying the current sweep rate,
  $\varkappa$.}
\end{thebibliography}

%====================================================================================================================================================%

\end{document}